\documentclass[aps,prl,twocolumn,superscriptaddress,showpacs,preprintnumbers,amsmath,amssymb]{revtex4}
\usepackage{epstopdf}
\usepackage{graphicx} % Include figure files
\usepackage{dcolumn}  % Align table columns on decimal point

\graphicspath{{ps}}

\begin{document}

%\vspace*{-3\baselineskip}
%\resizebox{!}{3cm}{\includegraphics{belle.eps}}

\preprint{\vbox{ \hbox{   }
%						\hbox{Belle DRAFT {\it v9.2}}
%                        \hbox{Intended for {\it PRD(RC)}}
%                        \hbox{Author: M.-C Chang, Y.-C. Duh, J.-Y. Lin}
%                        \hbox{Committee: Chung-Hsiang Wang (chair),}
%                        \hbox{Mizuki Sumihama, Gocha Tatishvili, }
                          \hbox{Belle Preprint 2012-10}
                          \hbox{KEK Preprint 2012-1}
%  		              % \hbox{hep-ex nnnn}
}}

\title{ \quad\\[1.0cm] Measurement of  $B^0 \to J/\psi \eta^{(}{}'{}^{)}$ and Constraint on the $\eta-\eta'$ Mixing Angle  }

%%%% >>>>> insert the authorlist here. BEFORE the abstract !!!!! <<<<<
%%%% >>>>> from the authorship confirmation web page
%%% Name the file author.tex and use \input{author} to insert into your latex file.
%\author{M.-C. Chang}\affiliation{Fu Jen Catholic University}
%\collaboration{The Belle Collaboration}
%\noaffiliation
%% end author list
%%% Paper:    B -> J/psi eta(')
%%% Journal:  Physical Review D (Rapid Communication)
%%% Contacts: M.C. Chang (fj068190@mail.fju.edu.tw)
%%% Non-responding authors or those who said NO are commented out.
%%% ====================================================================
%%% Click the RELOAD button on your web browser to see the updated file.
%%% ====================================================================
%%% Use \input{author} to insert this material into your latex file.
%%%%% Force institutions to appear in alphabetical order when typeset.
\affiliation{University of Bonn, Bonn}
\affiliation{Budker Institute of Nuclear Physics SB RAS and Novosibirsk State University, Novosibirsk 630090}
\affiliation{Faculty of Mathematics and Physics, Charles University, Prague}
%%%\affiliation{Chiba University, Chiba}
\affiliation{University of Cincinnati, Cincinnati, Ohio 45221}
\affiliation{Department of Physics, Fu Jen Catholic University, Taipei}
\affiliation{Justus-Liebig-Universit\"at Gie\ss{}en, Gie\ss{}en}
\affiliation{Gifu University, Gifu}
%%%\affiliation{The Graduate University for Advanced Studies, Hayama}
\affiliation{Gyeongsang National University, Chinju}
\affiliation{Hanyang University, Seoul}
\affiliation{University of Hawaii, Honolulu, Hawaii 96822}
\affiliation{High Energy Accelerator Research Organization (KEK), Tsukuba}
\affiliation{Hiroshima Institute of Technology, Hiroshima}
%%%\affiliation{University of Illinois at Urbana-Champaign, Urbana, Illinois 61801}
\affiliation{Indian Institute of Technology Guwahati, Guwahati}
\affiliation{Indian Institute of Technology Madras, Madras}
%%%\affiliation{Indiana University, Bloomington, Indiana 47408}
\affiliation{Institute of High Energy Physics, Chinese Academy of Sciences, Beijing}
\affiliation{Institute of High Energy Physics, Vienna}
\affiliation{Institute of High Energy Physics, Protvino}
%%%\affiliation{Institute of Mathematical Sciences, Chennai}
%%%\affiliation{INFN - Sezione di Torino, Torino}
\affiliation{Institute for Theoretical and Experimental Physics, Moscow}
\affiliation{J. Stefan Institute, Ljubljana}
\affiliation{Kanagawa University, Yokohama}
\affiliation{Institut f\"ur Experimentelle Kernphysik, Karlsruher Institut f\"ur Technologie, Karlsruhe}
\affiliation{Korea Institute of Science and Technology Information, Daejeon}
\affiliation{Korea University, Seoul}
%%%\affiliation{Kyoto University, Kyoto}
\affiliation{Kyungpook National University, Taegu}
\affiliation{\'Ecole Polytechnique F\'ed\'erale de Lausanne (EPFL), Lausanne}
\affiliation{Faculty of Mathematics and Physics, University of Ljubljana, Ljubljana}
\affiliation{Luther College, Decorah, Iowa 52101}
\affiliation{University of Maribor, Maribor}
\affiliation{Max-Planck-Institut f\"ur Physik, M\"unchen}
\affiliation{University of Melbourne, School of Physics, Victoria 3010}
\affiliation{Graduate School of Science, Nagoya University, Nagoya}
\affiliation{Kobayashi-Maskawa Institute, Nagoya University, Nagoya}
%%%\affiliation{Nara University of Education, Nara}
\affiliation{Nara Women's University, Nara}
\affiliation{National Central University, Chung-li}
\affiliation{National United University, Miao Li}
\affiliation{Department of Physics, National Taiwan University, Taipei}
\affiliation{H. Niewodniczanski Institute of Nuclear Physics, Krakow}
\affiliation{Nippon Dental University, Niigata}
\affiliation{Niigata University, Niigata}
\affiliation{University of Nova Gorica, Nova Gorica}
\affiliation{Osaka City University, Osaka}
%%%\affiliation{Osaka University, Osaka}
\affiliation{Pacific Northwest National Laboratory, Richland, Washington 99352}
\affiliation{Panjab University, Chandigarh}
%%%\affiliation{Peking University, Beijing}
%%%\affiliation{Princeton University, Princeton, New Jersey 08544}
\affiliation{Research Center for Nuclear Physics, Osaka University, Osaka}
%%%\affiliation{RIKEN BNL Research Center, Upton, New York 11973}
%%%\affiliation{Saga University, Saga}
\affiliation{University of Science and Technology of China, Hefei}
\affiliation{Seoul National University, Seoul}
%%%\affiliation{Shinshu University, Nagano}
\affiliation{Sungkyunkwan University, Suwon}
\affiliation{School of Physics, University of Sydney, NSW 2006}
\affiliation{Tata Institute of Fundamental Research, Mumbai}
\affiliation{Excellence Cluster Universe, Technische Universit\"at M\"unchen, Garching}
\affiliation{Toho University, Funabashi}
\affiliation{Tohoku Gakuin University, Tagajo}
\affiliation{Tohoku University, Sendai}
\affiliation{Department of Physics, University of Tokyo, Tokyo}
\affiliation{Tokyo Institute of Technology, Tokyo}
\affiliation{Tokyo Metropolitan University, Tokyo}
\affiliation{Tokyo University of Agriculture and Technology, Tokyo}
%%%\affiliation{Toyama National College of Maritime Technology, Toyama}
\affiliation{CNP, Virginia Polytechnic Institute and State University, Blacksburg, Virginia 24061}
%%%\affiliation{Wayne State University, Detroit, Michigan 48202}
\affiliation{Yamagata University, Yamagata}
\affiliation{Yonsei University, Seoul}
  \author{M.-C.~Chang}\affiliation{Department of Physics, Fu Jen Catholic University, Taipei} % FuJen
  \author{Y.-C. Duh}\affiliation{Department of Physics, Fu Jen Catholic University, Taipei} % FuJen
  \author{J.-Y. Lin}\affiliation{Department of Physics, Fu Jen Catholic University, Taipei} % FuJen
  \author{I.~Adachi}\affiliation{High Energy Accelerator Research Organization (KEK), Tsukuba} % KEK
  \author{K.~Adamczyk}\affiliation{H. Niewodniczanski Institute of Nuclear Physics, Krakow} % Krakow
  \author{H.~Aihara}\affiliation{Department of Physics, University of Tokyo, Tokyo} % Tokyo
% \author{K.~Arinstein}\affiliation{Budker Institute of Nuclear Physics SB RAS and Novosibirsk State University, Novosibirsk 630090} % BINP
% \author{Y.~Arita}\affiliation{Graduate School of Science, Nagoya University, Nagoya} % Nagoya
  \author{D.~M.~Asner}\affiliation{Pacific Northwest National Laboratory, Richland, Washington 99352} % PNNL
% \author{T.~Aso}\affiliation{Toyama National College of Maritime Technology, Toyama} % Toyama
% \author{V.~Aulchenko}\affiliation{Budker Institute of Nuclear Physics SB RAS and Novosibirsk State University, Novosibirsk 630090} % BINP
  \author{T.~Aushev}\affiliation{Institute for Theoretical and Experimental Physics, Moscow} % ITEP
% \author{T.~Aziz}\affiliation{Tata Institute of Fundamental Research, Mumbai} % Tata
  \author{A.~M.~Bakich}\affiliation{School of Physics, University of Sydney, NSW 2006} % Sydney
% \author{Y.~Ban}\affiliation{Peking University, Beijing} % Peking
% \author{E.~Barberio}\affiliation{University of Melbourne, School of Physics, Victoria 3010} % Melbourne
% \author{A.~Bay}\affiliation{\'Ecole Polytechnique F\'ed\'erale de Lausanne (EPFL), Lausanne} % Lausanne
% \author{I.~Bedny}\affiliation{Budker Institute of Nuclear Physics SB RAS and Novosibirsk State University, Novosibirsk 630090} % BINP
% \author{M.~Belhorn}\affiliation{University of Cincinnati, Cincinnati, Ohio 45221} % Cincinnati
% \author{K.~Belous}\affiliation{Institute of High Energy Physics, Protvino} % Protvino
  \author{V.~Bhardwaj}\affiliation{Nara Women's University, Nara} % Nara
% ...
% Karlsruhe
% \author{N.~Zwahlen}\affiliation{\'Ecole Polytechnique F\'ed\'erale de Lausanne (EPFL), Lausanne} % Lausanne
% \author{O.~Zyukova}\affiliation{Budker Institute of Nuclear Physics SB RAS and Novosibirsk State University, Novosibirsk 630090} % BINP
\collaboration{The Belle Collaboration}

\begin{abstract}
We measure the branching fractions of  $B^{0} \to J/\psi \eta^{(}{}'{}^{)}$ decays 
with the complete Belle data sample of $772 \times 10^{6}$ 
$B\bar{B}$ events collected at the $\Upsilon(4S)$ resonance with the Belle detector at the KEKB
asymmetric-energy $e^+ e^-$ collider.
The results for  the branching fractions are: 
${\cal B}(B^{0} \to J/\psi \eta)=(12.3 \pm ^{1.8} _{1.7} \pm 0.7) \times 10^{-6}$ and
${\cal B}(B^{0} \to J/\psi\eta') < 7.4 \times 10^{-6}$ at 90\% confidence level.
The $\eta-\eta'$ mixing angle is constrained to be less than
$ 42.2^{\circ}$ at 90\% confidence level.

\end{abstract}


\maketitle

%%%% >>>> keep the final version single-spaced
\tighten

{\renewcommand{\thefootnote}{\fnsymbol{footnote}}}
\setcounter{footnote}{0}

\begin{figure}
\centering
\includegraphics[width=0.3\textwidth]{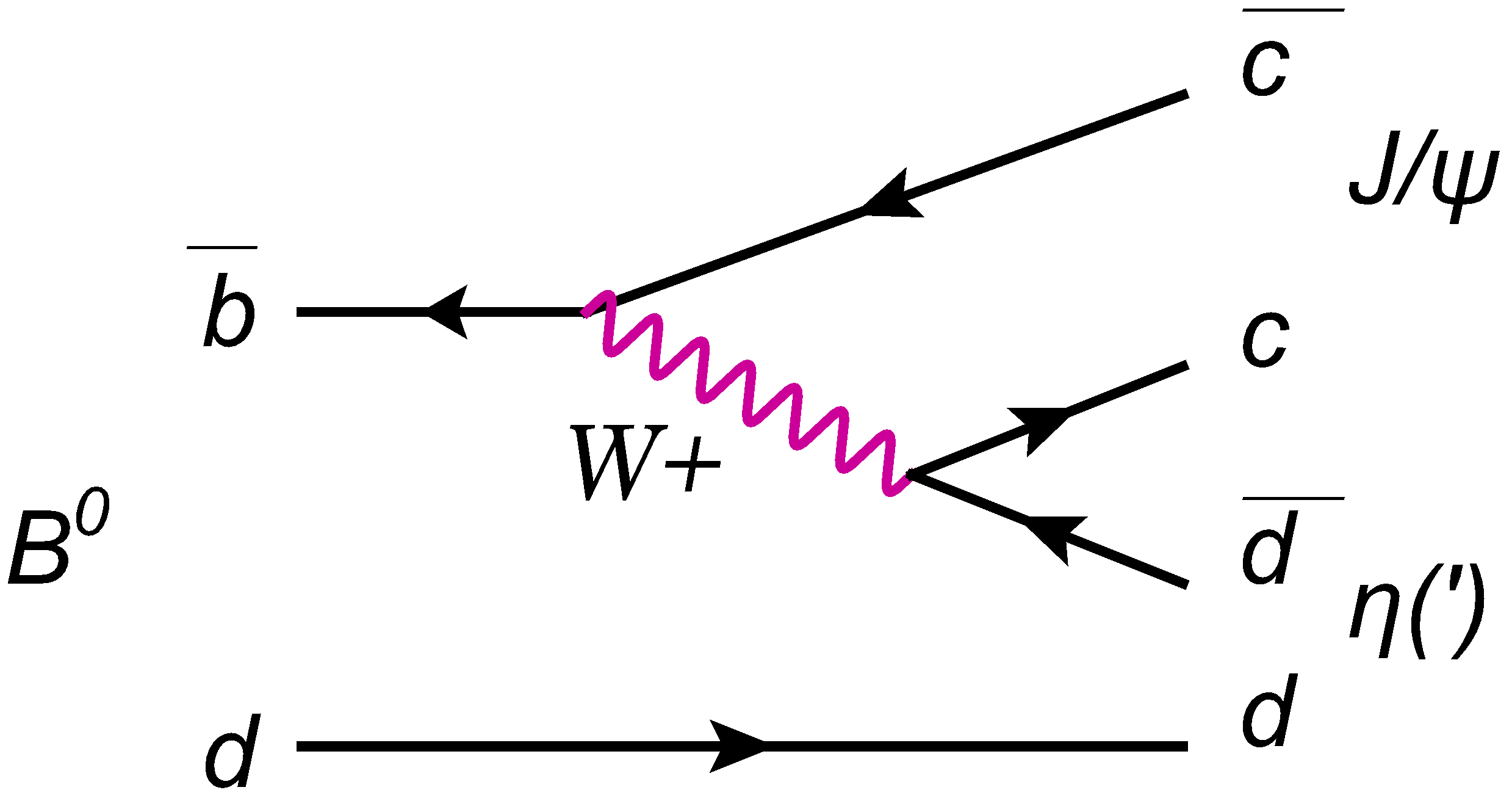} 
\caption{Quark level diagram of the leading transition  
$ B^0 \to J/\psi \eta^{(}{}'{}^{)}$. }~\label{feynman_plot}
\end{figure}

The neutral $B$ meson decays $B^0 \to J /\psi \eta^{(}{}'{}^{)}$ are mediated by 
the $\bar{b} \to c\bar{c}\bar{d}$ transition as shown in Fig.~\ref{feynman_plot}.  
For such final states involving an $\eta$ or $\eta'$ meson, it is convenient to consider
 flavor mixing of $\eta_q$ and $\eta_s$ defined by
\begin{align}~\label{step1}
\eta_{q}=\dfrac{1}{\sqrt{2}}(u \bar{u}+d \bar{d}), \
\eta_{s}=s \bar{s},
\end{align} 
in analogy with the wave-functions of $\omega$ and $\phi$ for  ideal mixing~\cite{PQCD1}.
The wave-functions of the $\eta$ and $\eta'$ are given by 
 \begin{equation}~\label{step2}
\left(   
\begin{array}{c}
  \eta  \\
  \eta'  
\end{array}\right)
=
\left(   
\begin{array}{cc}
  \cos \phi  &-\sin \phi\\
  \sin \phi  & \cos \phi 
\end{array}\right)
\left(   
\begin{array}{c}
  \eta_q  \\
  \eta_s 
\end{array}\right).
\end{equation}
%where $\phi=\theta+$arctan$\sqrt{2}$, and $\theta$ is the $\eta-\eta'$ mixing angle in the octet-singlet basis 
%\begin{equation}
%\left(   
%\begin{array}{c}
 % \eta  \\
 % \eta'  
%\end{array}\right)
%=
%\left(   
%\begin{array}{cc}
%  \cos \theta  &-\sin \theta\\
%  \sin \theta  & \cos \theta 
%\end{array}\right)
%\left(   
%\begin{array}{c}
%  \eta_8  \\
%  \eta_0 
%\end{array}\right).
%\end{equation}
If the  $s\bar{s}$ component in Eq.~(\ref{step2}) for
$B^0 \to J/\psi \eta^{(}{}'{}^{)}$ decays is negligible,
the branching fractions are related to the 
$\eta-\eta'$ mixing angle $\phi$ as: 
\begin{align}~\label{mixing}
\dfrac{\cal{B}\rm{(B^0 \to J/\psi \eta')}}{\cal{B}\rm{( B^0 \to J/\psi \eta)}} \simeq \tan^2 \phi .
\end{align} 
Using the measured values of the $\eta -\eta'$ mixing angle 
$\phi \sim  40^\circ$ ~\cite{KLOE} and 
${\cal B}(B^0 \to J /\psi \eta) = (9.5 \pm 1.7 \pm 0.8) 
\times 10^{-6} $~\cite{PRLjpsieta} in Eq.~(\ref{mixing}), 
the expected branching fraction 
for $B^0 \to J /\psi \eta'$ is about $6.7 \times 10^{-6} $.
Other predictions
give ${\cal B}(B^0 \to J /\psi \eta')=(4-8) \times 10^{-6}$ and  
$\eta -\eta'$ mixing angle $37^{\circ}<\phi<50^{\circ}$
~\cite{PQCD2,PQCD3,PQCD4,PQCD5}.
Recently, $B_s^0 \to J/\psi \eta^{(}{}'{}^{)}$ decays have been observed by Belle~\cite{Li-Jin}. The branching fractions in Ref.~\cite{Li-Jin} and this paper provide good inputs for  model predictions~\cite{PQCD2,PQCD3,PQCD4,PQCD5}.
The existing upper limit for the $B^0 \to J /\psi \eta'$ branching fraction is $6.3 \times 10^{-5} $~\cite{babar}.

In this paper, we report a measurement of  $B^0 \to J /\psi \eta$ 
and a search for $B^0 \to J /\psi \eta'$ decays~\cite{CC}.
The results are based on a data sample that contains 
$772 \times 10^{6}$ $B\overline{B}$ pairs, collected with the Belle detector 
at the KEKB asymmetric-energy $e^+e^-$ (3.5 on 8 GeV) collider~\cite{KEKB} 
operating at the $\Upsilon (4S)$ resonance.

The Belle detector is a large-solid-angle magnetic
spectrometer that consists of a silicon vertex detector (SVD),
a 50-layer central drift chamber (CDC), an array of
aerogel threshold Cherenkov counters (ACC),  % <- \v{C}erenkov 2007.08
a barrel-like arrangement of time-of-flight
scintillation counters (TOF), and an electromagnetic calorimeter
 (ECL) located inside 
a superconducting solenoid coil that provides a 1.5~T
magnetic field.  An iron flux-return located outside of
the coil is instrumented to detect $K_L^0$ mesons and to identify
muons (KLM).  The detector
is described in detail elsewhere~\cite{Belle}.
% {\bf SVD2+SVD1, up to experiment 37:}

The data sample used in this analysis was collected with two detector 
configurations. A 2.0 cm beampipe and a 3-layer silicon vertex detector 
were used for the first sample
of $152 \times 10^6 B\bar{B}$ pairs, while a 1.5 cm beampipe, a 4-layer
silicon detector and a small-cell inner drift chamber were used to record  
the remaining $620 \times 10^6 B\bar{B}$ pairs~\cite{svd2}. 
GEANT-based Monte-Carlo(MC) simulation program is used to model the 
   response of the detector and to determine the reconstruction 
   efficiencies.  Simulated events are generated with the EvtGen program~\cite{evtgen}, except for the $B^0 \to J /\psi \eta'$($\eta' \to \rho^0 \gamma$) 
   mode for which we use the QQ program~\cite{qq98} since EvtGen does not 
  properly generates the $\rho^0$ mass and angular distributions.

Charged tracks are selected by using the impact parameters relative 
to the interaction point: $dr$ for the radial direction and $dz$ 
for the direction along the positron beam. Our requirements are 
$dr < 1 $ cm and $ |dz| < 5 $ cm. Identification of $e^{+}$ and $e^{-}$ 
from $J/\psi$ decay  uses information from the ECL, 
the CDC $(dE/dx)$ and the ACC. Identification of $ \mu^+$ and $ \mu^-$ 
candidates uses a track penetration depth and hit pattern in 
the KLM system. Charged pions are identified based on the information from the CDC ($dE/dx$), the TOF and the ACC.

Photon candidates are selected from showers in the ECL, 
which are not associated with charged tracks, and an energy deposition 
of at least 50 MeV in the barrel region or 100 MeV in the endcap region. A pair of photons with an 
invariant mass in the range $117.8~{\rm MeV}/c^2  < M_{\gamma\gamma} < 150.2~{\rm MeV}/c^2 $ 
is considered as a $ \pi^0 $ candidate. This invariant mass region 
corresponds to a $\pm$~3$\sigma$ interval around the $\pi^0$ mass, 
where $\sigma$ is the mass resolution.

We reconstruct $J /\psi$ mesons in the $ l^+l^- $ decay channel 
($ l = e$ or $\mu $).  
Any  photon within 50 mrad  of $e^+$ or $e^-$ tracks  is included as well. The invariant mass 
is required to be within $-0.15~{\rm GeV}/c^2 < M_{ee(\gamma)} - m_{J /\psi} < 0.036~ {\rm GeV}/c^2$ and 
$-0.06~{\rm GeV}/c^2  < M_{\mu\mu}  - m_{J /\psi} < 0.036~{\rm GeV}/c^2$, 
where $m_{J/ \psi}$ denotes the nominal $J/\psi$ mass~\cite{PDG}, $M_{ee(\gamma)}$ and 
$M_{\mu\mu}$ are the reconstructed invariant mass of $e^+ e^- (\gamma)$ 
and $\mu^+ \mu^-$, respectively. An asymmetric interval is used to 
include part of the radiative tails.

$\eta$ mesons are reconstructed in the $\gamma \gamma$ and 
$\pi^+ \pi^- \pi^0$ final states. 
The mass ranges are $497~{\rm MeV}/c^2 < M_{\gamma\gamma} < 590~{\rm MeV}/c^2$  and 
 $520~{\rm MeV}/c^2 < M_{\pi^+\pi^-\pi^0} < 557~{\rm MeV}/c^2$.
 In the $ \gamma \gamma $ final state, candidates in which 
either of the daughter photons forms a $\pi^0$ together with any other 
photon in the event are rejected.  In the 
$ \gamma \gamma $ final state, we require $ | \cos \theta_{\eta} | < 0.9 $, 
where $\theta_{\eta}$ is defined as the angle between the momentum of either of the photons and the boost direction of the laboratory system in the rest frame of the $\eta$.

$\eta'$ mesons are reconstructed in the $ \eta \pi^{+} \pi^{-}$ ($\eta \to \gamma \gamma$) and 
$\rho^{0} \gamma$ final states. 
The mass ranges are 
 $930~{\rm MeV}/c^{2} < M_{\eta\pi^{+}\pi^{-}} < 967~{\rm MeV}/c^{2}$, 
$920~{\rm MeV}/c^{2} < M_{\rho^{0}\gamma} < 980~{\rm MeV}/c^{2}$, and 
$600~{\rm MeV}/c^{2} < M_{\rho^{0}} < 900~{\rm MeV}/c^{2}$. 
 In the  $\rho^{0} \gamma$ final state, we require 
 $  \cos \theta_{\eta'}  < 0.6 $  and $  |\cos \theta_{\rho}|  < 0.8 $. 
Here $\theta_{\eta'}$ is the angle   between the  photon momentum and the opposite of the boost direction of the laboratory system in the $\eta'$ rest frame. 
The other angle, $\theta_{\rho}$, is the angle   between the $\pi^+$ momentum and 
 the boost direction of the laboratory system in the $\rho$ rest frame.

$B$ mesons are reconstructed in the $J/\psi\eta^{(}{}'{}^{)}$ final state. 
Signal candidates are identified using two kinematic variables defined 
in the $\Upsilon (4S)$ center-of-mass (CM) frame: the beam-energy 
constrained mass, $M_{\rm bc}$ = $\sqrt{E^{2}_{\rm beam} - p^{*2}_{\rm B}}$ and the 
energy difference, $ \Delta E = E^{*}_{\rm B} - E_{\rm beam}$, where $p^{*}_{\rm B}$ and 
$E^{*}_{\rm B}$ are the momentum and energy of the $B$ candidate and $E_{\rm beam}$ 
is the run-dependent beam energy. To improve the momentum resolution, 
the masses of the selected $\pi^0$, $\eta^{(}{}'{}^{)}$ and $J /\psi$ candidates 
are constrained to their nominal masses using mass-constrained kinematic
fits. In addition, vertex-constrained fits are applied to 
$\eta \to \pi^+ \pi^- \pi^0$, $\eta' \to \eta \pi^+ \pi^- $ and 
$J /\psi \to l^+ l^-$ candidates. 
We  retain  events with $M_{\rm bc} > 5.2~ {\rm GeV}/c^2$ and $|\Delta E| < 0.2$ GeV. 
The signal peaks in the region defined by 5.27 ${\rm GeV}/c^2 < M_{\rm bc} < 5.29~{\rm GeV}/c^2$, 
$|\Delta E| < 0.05~{\rm GeV}$ 
(or $-0.10~{\rm GeV} <  \Delta E < 0.05$~GeV 
 for $B^0\to J/\psi \eta(\eta \to \gamma \gamma)$ only).

For events with more than one $B$ candidate, which are usually due to multiple $\eta^{(}{}'{}^{)}$ candidates,  the candidate with
 the minimum $\chi^{2}$ value from the mass- and vertex-constrained fit is chosen.

The combinatorial background dominated by 
two-jet-like $e^+ e^- \to q\overline{q} (q = u,d,s,c)$ continuum 
is suppressed by requiring the ratio of second to zeroth 
Fox-Wolfram moments $R_2 < 0.4$~\cite{SFW}. 

After the continuum suppression, the background is dominated 
by $B\overline{B}$ events with $B \to J /\psi X$ decays, where $X$ denotes any final state. 
Using an MC sample of generic $B\overline{B}$ decays corresponding to 10 times larger than the data, 
with   all known and expected $B^0 \to J /\psi X$ decays, we study these backgrounds. 
The backgrounds from
$B^0 \to J /\psi K$  and $B^0 \to J /\psi \pi^0$ cannot be highly suppressed  and
 peak in the $M_{bc}$ distributions but not in $\Delta E$. 

A figure-of-merit (FOM) method is used to optimize the selection requirements. 
The maximal value of  $N_{s}/\sqrt{N_{s} + N_{b}}$ is chosen for each variable, where $N_{s}$ is the number of expected 
signal events and $N_{b}$ is the number of background events. 

For each requirement, $N_{s}$ is estimated from signal MC simulation as
\begin{align}~\label{ns}
N_s=N_{B\bar{B}}{\cal B}(B^0 \to J/\psi \eta')\varepsilon {\cal B}_{sec},
\end{align}
where   
$N_{B\bar{B}}$ is the total number of $B\bar{B}$ pairs,
$6.7 \times 10^{-6}$ is assumed for ${\cal B}(B^0 \to J /\psi \eta')$,
 $\varepsilon$
is the signal efficiency, and ${\cal B}_{sec}$ is a product of the 
branching fractions for secondary decays.

The value of $N_b$ is estimated from
\begin{align}~\label{nb}
N_b = R_{MC}  N_{b}^{'},
\end{align}
where $R_{MC}$ is the  
fraction of $B \to J/\psi X$ MC events in the signal region, 
and  $N_{b}^{'}$ is the number of data events in the $\Delta E$ sideband region.

Signal yields and background levels are determined by fitting 
the $\Delta E$ distribution  for candidates in the $M_{bc}$ signal region. 
For $B^0 \to J /\psi \eta$, the $\Delta E$ distribution is fitted using a
 signal probability density function (PDF),  which is a sum of a 
Crystal Ball function~\cite{CB} and two Gaussian functions. For $B^0 \to J /\psi \eta'$, 
the signal PDF is a sum of three Gaussian functions. 
%We denote the former signal PDF as PDF$_{s1}$ and the latter as PDF$_{s2}$.  
The background PDF is a second-order polynomial function. 
The signal shapes are determined from  MC samples. 
The background levels are floated. 

We use the $B^+ \to J /\psi K^{*+}(K^{*+} \to K^{+} \pi^0)$ decay as a 
control sample to correct the difference between data and MC 
in the fitted mean and width of the $\Delta E$ signal peak. 
We require the helicity angle $\theta_{K^*}$ in the $K^{*+} \to K^+ \pi^0$ 
decay to be less than 90 degrees.  
Here $\theta_{K^*}$ is the angle between the $\pi^0$ momentum and the opposite of the boost direction of the laboratory system in the  $ K^{*+} $ rest frame.
This requirement primarily selects events with a high momentum $\pi^0$ 
and produces a control-sample $\Delta E$ distribution that is 
similar to that in our decay.

 The signal PDFs are modified based on the differences of mean and width between 
data and MC in the control sample. From a fit  to the control sample, we find that the mean values are 
shifted by $(-3.85~\pm~0.13)~\rm {MeV}$. 
The width  in data is $(1.11~\pm~0.03)$ times wider than in MC simulation.

%In the $B^0 \to J /\psi \eta$, there are 164 and 75 events in the $M_{bc}$ signal region for the $\eta \to \gamma\gamma$ and $\eta \to \pi^+ \pi^- \pi^0$ modes. And in the $B^0 \to J /\psi \eta'$ mode, there are 91 and 3324 events in the $M_{bc}$ signal region for the $\eta' \to \eta\gamma\gamma$ and $\eta' \to \rho^0\gamma$ modes, respectively. 

We determine the signal yields by performing an unbinned extended 
maximum-likelihood fit to the candidate data events:
\begin{align}
\mathcal{L}= \frac{e^{-(N_{s}+N_{b})}}{N} \prod_{i}^N{} 
[N_{s}P_{s}(\Delta E_{i})+N_{b}P_{b}(\Delta E_{i})].
\end{align}

Here $N$ is the total number of candidate events,  $P_{s}(\Delta E_{i})$ 
and $P_{b}(\Delta E_{i})$ denote the signal and background $\Delta E$ PDFs,
respectively, and $i$ is the event index. 

\begin{figure}
\centering
\includegraphics[width=0.5\textwidth]{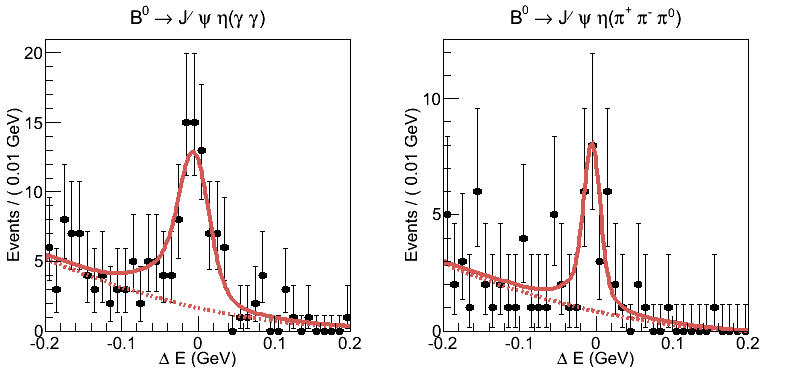}
\includegraphics[width=0.5\textwidth]{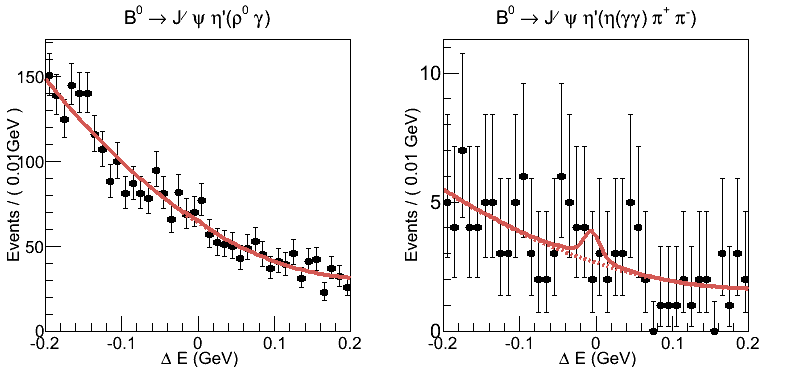} 
\caption{$\Delta E$ distributions for the two decay modes,  
$B^0 \to J /\psi \eta$ and $B^0 \to J /\psi \eta'$. 
The plots on the top are $B^0 \to J /\psi \eta$,  in the left side --
$B^0 \to J /\psi \eta(\eta \to \gamma\gamma)$, in the right side -- 
$B^0 \to J /\psi \eta(\eta \to \pi^{+}\pi^{-}\pi^{0})$. The plots on the bottom 
are $B^0 \to J /\psi \eta'$, in the  left side --
$B^0 \to J /\psi \eta'(\eta' \to \rho^{0}\gamma)$, in the right side -- 
$B^0 \to J /\psi \eta'(\eta' \to \eta\pi^{+}\pi^{-})$. The solid curves are 
results of the overall fit while the dashed curves are the 
background shapes.}~\label{data_plot}
\end{figure}

In the $B^0 \to J /\psi \eta$ mode, we fit the 
$B^0 \to J /\psi \eta(\gamma\gamma)$ 
and $B^0 \to J /\psi \eta(\pi^+ \pi^- \pi^0)$ candidate samples simultaneously with a common branching fraction.
The fit gives the branching faction $(12.3 \pm ^{1.8} _{1.7}) \times 10^{-6}$, which
   corresponds to  signal yields of 77.9 and 29.8 for the $\eta(\gamma\gamma)$
   and $\eta(\pi^+\pi^-\pi^0)$ modes, respectively.
For the $B^0 \to J /\psi \eta'$ mode, we also fit the 
$B^0 \to J /\psi \eta'(\eta\pi^{+}\pi^{-})$ and 
$B^0 \to J /\psi \eta'(\rho^{0}\gamma)$ candidate samples simultaneously with a  common branching fraction.
The fit gives the branching fraction $(2.2 \pm ^{3.3} _{2.9}) \times 10^{-6}$, which
   corresponds to  signal yields of 
5.5 and 5.2 for the $\eta'(\rho^{0}\gamma)$
   and  $\eta'(\eta\pi^{+}\pi^{-})$ modes, respectively.
The results of the fits to the data are shown in Fig.~\ref{data_plot}.

%A X.X\% systematic error associated with uncertainties in the signal and background PDFs is estimated by comparing the fit results for the cases when the polynomial parameters are fixed by either MC or data in the $M_{bc}$ sideband region, and from changes that result from varying each parameter by one standard deviation.

A 1.4\% systematic error comes from the uncertainty in 
the number of $B\bar{B}$ pairs.
The systematic error due to tracking  is 0.35\% for each charged track.
The systematic error from the pion identification requirement is determined 
from a study of the $D^{*+}\to D^0 \pi^+(D^0 \to K^- \pi^+)$ control sample. 
The systematic error from lepton identification is obtained from a 
comparison between the data and  MC for $\gamma \gamma \to e^+e^-/\mu^+\mu^-$ 
events.

A 3.0\% systematic error due to $\pi^0$ detection is determined from a 
comparison of the data and MC ratios for a large sample of 
$\eta \to \pi^+ \pi^- \pi^0$ and $\eta \to 3\pi^0$ decays. 
Since $\eta \to \gamma \gamma$ is similar to $\pi^0$ decay, 
we also assign a 3.0\%  systematic error for $\eta \to \gamma \gamma$ 
reconstruction.

We use the previously mentioned control sample 
$B^+ \to J /\psi K^{*+}(K^{*+} \to K^{+} \pi^0)$ to obtain the systematic 
errors from the $R_2<0.4$ requirement and signal PDFs. A 3.3\% systematic error 
 is obtained by comparing the data and MC. 
The systematic errors from signal PDFs are obtained by comparing the 
fit results for the cases when the fitting parameters are fixed from either 
MC or data and from changes that result from varying each parameter by 
one standard deviation. 
The errors on the branching fractions are taken from Ref.~\cite{PDG}.
%The errors on the branching fractions for 
%${\cal B}(J/\psi \to \l^{+}\l^{-})$, ${\cal B}(\eta \to \gamma\gamma)$, 
%${\cal B}(\eta \to \pi^{+}\pi^{-}\pi^{0}$), 
%${\cal B}(\eta' \to \eta\pi^{+}\pi^{-}$), and ${\cal B}(\eta' \to \rho^{0}\gamma)$ are
%1.0\%, 0.5\%, 1.2\%, 1.6\%, and 2.0\%, respectively~\cite{PDG}.

The systematic errors are summarized in Tables~\ref{sys1} and~\ref{sys2}. The total 
uncertainty is calculated by summing all individual uncertainties in quadrature. 

The statistical significances of the observed 
$B^0 \to J /\psi \eta$ and $B^0 \to J /\psi \eta'$ yields 
are 6.3$\sigma$ and 0.4$\sigma$, respectively. The significance is defined as 
$\sqrt{-2{\rm ln}(L_{0}/L_{\rm max})}$, where $L_{\rm max}(L_{0})$ denotes the 
likelihood value at the maximum (with the signal yield fixed at zero). 
The background is floated and hence  the systematic error is included in the significance calculation. 

An  upper limit has been evaluated for the branching fraction of 
$B^0 \to J /\psi \eta'$ at 90\% confidence level because of the low 
significance. 
The Bayesian upper limits are obtained from:
\begin{align}
\int^{N}_{0}L(n)dn=0.9\int^{\infty}_{0}L(n)dn,
\end{align}
where $N$ denotes  the signal yield.
%We  modify the likelihood function using  a smeared likelihood function: 
We use a modified likelihood function to obtain a conservative upper limit,
which is smeared with the systematic errors in the branching fraction.
%\begin{align}~\label{s-likelihood}
%L_{\rm smear}(N_{\rm sig})=\int L(N_{\rm sig}')\dfrac{e^{-\dfrac{(N_{\rm sig}-N_{\rm sig}')^{2}}{2\Delta %N_{\rm sig}^{2}}}}{\sqrt{2\pi \Delta N_{\rm sig}}}dN_{\rm sig}'
%\end{align}
The upper limit for the branching fraction of 
$B^0 \to J /\psi \eta'$ at 90\% confidence level is $ 7.4 \times 10^{-6}$.

\begin{table}[htb]
\caption{ Systematic uncertainties for $B \to J/\psi\eta$ (\%). The  combined systematic error is $5.9\%$.}
\label{sys1}
\begin{tabular}[t]{lcc}
\hline \hline
%{@{\hspace{0.5cm}}l@{\hspace{0.5cm}}||@{\hspace{0.5cm}}c@{\hspace{0.5cm}}}
%{@{\hspace{0.5cm}}l@{\hspace{0.5cm}}  @{\hspace{0.5cm}}c@{\hspace{0.5cm}}}
$\eta$ source &  $\gamma\gamma$  &  $\pi^{+}\pi^{-}\pi^{0}$   \\   
\hline
Number of $B\overline{B}$ events & 1.4 &  1.4  \\
Tracking &  0.7  &  1.4  \\
Lepton-ID  & 2.6 &   2.6  \\
Charged $\pi$-ID  &  -  &   2.4  \\
$\eta \to \gamma\gamma$, $\pi^0$ selection  & 3.0  &  3.0  \\
PDFs   &   1.3  &  1.3   \\
$R_2$ $<$ 0.4  &  3.3   &   3.3    \\
${\cal B}(J/\psi \to e^{+}e^{-},\mu^{+}\mu^{-})$  &   1.0   &   1.0  \\
${\cal B}(\eta \to \gamma\gamma)$   &   0.5  &     -    \\
${\cal B}(\eta \to \pi^{+}\pi^{-}\pi^{0})$      &   -    &      1.2       \\  
\hline
Total & 5.7 &  6.4 \\
\hline \hline

\end{tabular}
\end{table}

\begin{table}[htb]
\caption{ Systematic uncertainties for $B \to J/\psi\eta'$ (\%). 
 The  combined systematic 
error is $7.5\%$.}
\label{sys2}
\begin{tabular}[t]{lcc}
\hline \hline
%{@{\hspace{0.5cm}}l@{\hspace{0.5cm}}||@{\hspace{0.5cm}}c@{\hspace{0.5cm}}}
%{@{\hspace{0.5cm}}l@{\hspace{0.5cm}}  @{\hspace{0.5cm}}c@{\hspace{0.5cm}}}
$\eta'$ source &  $\eta\pi^{+}\pi^{-}$   &  $\rho^{0}\gamma$  \\   
\hline
Number of $B\overline{B}$ events &  1.4 & 1.4  \\
Tracking(lepton and charged pion) & 1.4  &  1.4  \\
Lepton-ID  &  2.6 &   2.6  \\
Charged $\pi$-ID  &  3.0  &  3.0  \\
$\eta \to \gamma\gamma$, $\pi^0$ selection  & 3.0  &  3.0  \\
PDFs   &   3.6   &   3.6   \\
$R_2$ $<$ 0.4  &  3.3  &  3.3  \\
${\cal B}(J/\psi \to e^{+}e^{-},\mu^{+}\mu^{-})$  &   1.0   &   1.0  \\
${\cal B}(\eta \to \gamma\gamma)$   &   0.5  &   -   \\  
${\cal B}(\eta' \to \rho^{0}\gamma)$ &       -        &  2.0           \\
${\cal B}(\eta' \to \eta\pi^{+}\pi^{-})$  &   1.6      &  -         \\  
\hline
Total & 7.5 & 7.6 \\
\hline \hline

\end{tabular}
\end{table}

The signal efficiencies and the branching fractions 
are listed in Table~\ref{br1}.

\begin{table}[htb]
\caption{MC efficiencies and branching fractions.  }
\label{br1}
\begin{tabular}[t]{lcc}
\hline \hline
Mode  & MC Efficiency(\%)   &    ${\cal B}(10^{-6})$  \\
\hline
$B^{0} \to J/\psi \eta(\gamma\gamma)$  &  35.7  &     \\
$B^{0} \to J/\psi \eta(\pi^+\pi^-\pi^0)$ & 24.6  &     \\
$B^{0} \to J /\psi\eta$ combined        &        & $12.3 \pm ^{1.8} _{1.7} \pm 0.7 $  \\
\hline
$B^0 \to J/\psi \eta'(\eta\pi^+\pi^-$) & 31.0 &     \\
$B^{0} \to J /\psi \eta' (\rho^{0}\gamma$) & 19.1 &     \\
$B^0 \to J/\psi\eta'$ combined  & & $2.2 \pm ^{3.3} _{2.9} \pm 0.2$ \\
$B^0 \to J/\psi\eta'$ UL (90\%) & & $<7.4$\\
\hline \hline
\end{tabular}
\end{table}

%\begin{table}[htb]
%\caption{ Detection efficiencies and branching fraction.}
%\label{br2}
%\begin{tabular}[t]{lll}
%\hline \hline
%Mode  & Eff.(\%)   &    BF($10^{-6}$)  \\
%\hline
%$\eta' \to \eta\pi^{+}\pi^{-}$   &  31.0  &  3.3 $\pm$ 2.8 $\pm$ 0.3   \\
%$\eta' \to \rho^{0}\gamma$                     &  19.1  & 15.7 $\pm$ 11.0 $\pm$ 1.6   \\
%$B^{0} \to J /\psi\eta'$                       &        &  3.9 $\pm$ 2.8 $\pm$ 0.4   \\
%\hline
%\end{tabular}
%\end{table}

In summary, we measure the branching fraction ${\cal B}(B^0 \to J /\psi \eta)$ = 
$(12.3 \pm ^{1.8} _{1.7} \pm 0.7) \times 10^{-6}$. The first error is statistical 
and the second is systematic. This result is consistent with and supersedes
the previous Belle measurement~\cite{PRLjpsieta}.  
We do not observe a significant signal in $B^0 \to J /\psi \eta'$ 
and set the upper limit ${\cal B}(B^0 \to J /\psi \eta')$  $< 7.4 \times 10^{-6}$
 at 90\% confidence level, 
which is eight times more stringent than the previous result~\cite{babar}. From Eq. 
(\ref{mixing}) we calculate the $\eta-\eta'$ mixing angle, which is less than 
$ 42.2^{\circ}$ at 90\% confidence level. These results are consistent with 
the theoretical predictions of Refs.~\cite{PQCD2,PQCD3,PQCD4,PQCD5}.

%***** Acknowledgments *****
%----------- Long version, for most papers ----------- 
We thank the KEKB group for the excellent operation of the
accelerator; the KEK cryogenics group for the efficient
operation of the solenoid; and the KEK computer group,
the National Institute of Informatics, and the 
PNNL/EMSL computing group for valuable computing
and SINET4 network support.  We acknowledge support from
the Ministry of Education, Culture, Sports, Science, and
Technology (MEXT) of Japan, the Japan Society for the 
Promotion of Science (JSPS), and the Tau-Lepton Physics 
Research Center of Nagoya University; 
the Australian Research Council and the Australian 
Department of Industry, Innovation, Science and Research;
the National Natural Science Foundation of China under
contract No.~10575109, 10775142, 10875115 and 10825524; 
the Ministry of Education, Youth and Sports of the Czech 
Republic under contract No.~LA10033 and MSM0021620859;
the Department of Science and Technology of India; 
the Istituto Nazionale di Fisica Nucleare of Italy; 
the BK21 and WCU program of the Ministry Education Science and
Technology, National Research Foundation of Korea,
and GSDC of the Korea Institute of Science and Technology Information;
the Polish Ministry of Science and Higher Education;
the Ministry of Education and Science of the Russian
Federation and the Russian Federal Agency for Atomic Energy;
the Slovenian Research Agency;  the Swiss
National Science Foundation; the National Science Council
and the Ministry of Education of Taiwan; and the U.S.\
Department of Energy and the National Science Foundation.
This work is supported by a Grant-in-Aid from MEXT for 
Science Research in a Priority Area (``New Development of 
Flavor Physics''), and from JSPS for Creative Scientific 
Research (``Evolution of Tau-lepton Physics'').

%***** Acknowledgments *****

%\begin{verbatim}
% http://belle.kek.jp/secured/publication/ack.txt
%\end{verbatim}

%


\begin{thebibliography}{99}
\bibitem{CC}
Throughout this paper, 
the inclusion of the charge-conjugate decay mode is implied
unless otherwise stated.
\end{thebibliography}
\end{document}